\documentclass[conference]{IEEEtran}
\usepackage{amsmath}
\usepackage{amsfonts}
\usepackage{amssymb}
\usepackage{makeidx}
\usepackage{graphicx}
\usepackage{epstopdf}
\usepackage{cite}
\usepackage{paralist}

\ifCLASSINFOpdf

 \newtheorem{theorem}{Theorem}

 \newtheorem{proof}{Proof}
 
\else

\fi

\newcommand{\yu}{\mathbf{y}_{\text{u}}}
\newcommand{\pu}{P_{\text{u}}}
\newcommand{\pr}{P_{\text{r}}}
\newcommand{\pp}{P_{\text{p}}}

\newcommand{\np}{\mathbf{N}_{\text{p}}}

\newcommand{\G}{\mathbf{G}}
\newcommand{\Gh}{\hat{\mathbf{G}}}
\newcommand{\g}{\mathbf{g}}
\newcommand{\gh}{\hat{\mathbf{g}}}
\newcommand{\e}{\mathbf{e}}
\newcommand{\ai}{{\tt a}}
\newcommand{\bi}{{\tt b}}
\newcommand{\ci}{{\tt c}}
\newcommand{\yr}{\mathbf{y}_{\text{R}}}
\newcommand{\yp}{\mathbf{Y}_{\text{P}}}
\newcommand{\B}[1]{\mathbf{#1}}

\newcommand{\CN}{\mathcal{CN}}

\begin{document}

\title{Multi-Way Massive MIMO with Maximum-Ratio Processing and Imperfect CSI}

\author{\IEEEauthorblockN{Chung Duc Ho\IEEEauthorrefmark{1},
Hien Quoc Ngo\IEEEauthorrefmark{01}\IEEEauthorrefmark{2},
Michail Matthaiou\IEEEauthorrefmark{1}, and Trung Q. Duong\IEEEauthorrefmark{1}}
\IEEEauthorblockA{\IEEEauthorrefmark{1}School of Electronics, Electrical Engineering and Computer Science, Queen's University Belfast, BT7 1NN, Belfast, U.K.}
\IEEEauthorblockA{\IEEEauthorrefmark{2}Department of Electrical Engineering (ISY),
Link\"{o}ping University, 581 83 Link\"{o}ping, Sweden} 
\IEEEauthorblockA{\IEEEauthorrefmark{0}Email:\{choduc01, m.matthaiou, trung.q.duong\}@qub.ac.uk, hien.ngo@liu.se}}

\maketitle

\begin{abstract}

This paper considers a multi-way massive multiple-input multiple-output  relaying system, where single-antenna users exchange their information-bearing signals with the help of one relay station equipped with unconventionally many antennas. The relay first estimates the channels to all users through the pilot signals transmitted from them. Then, the relay uses maximum-ratio processing (i.e. maximum-ratio combining in the multiple-access phase and maximum-ratio transmission in the broadcast phase) to process the signals. A rigorous closed-form expression for the spectral efficiency is derived. The effects of the channel estimation error, the channel estimation overhead, the length of the training duration, and the randomness of the user locations are analyzed. We show that by deploying massive antenna arrays at the relay and simple maximum-ratio processing, we can  serve many users in the same time-frequency resource, while maintaining a given quality-of-service for each user.

\end{abstract}


\IEEEpeerreviewmaketitle

\section{Introduction}

During the past decades, relaying networks have attracted a great deal of research attention from both academia and industry. In multi-way relay networks, many users simultaneously exchange their bearing information among them via the assistance of a single sharing relay at the same time-frequency resource \cite{Gunduz:13:IT, AK:09:PIRM}. Multi-way relay networks provide spatial diversity, and hence, they can scale-up the spectral efficiency of the system without increasing the system complexity \cite{ALP:15:TWCOM, AP:14:PIRM, AK:09:PIRM, ATA:13:TCOM}. In \cite{ATA:13:TCOM}, the authors show that the spectral efficiency of multi-way relay networks is much higher than that of one-way or two-way networks for both perfect and imperfect channel state information (CSI). For all these reasons, there is a plethora of applications of multi-way relay networks, including wireless conference, power control in heterogeneous cellular networks, and low complexity data exchange between sensor nodes and data fusion centres in wireless sensor nodes.

Massive multiple-input multiple-output (MIMO), where a base station equipped with hundreds of antennas serves many active users in the same time-frequency resource, is considered as one of the key candidates for next-generation wireless systems \cite{NLM:13:TCOM, Rusek:2013:SPM,  EGL:14:CM}. There is a large body of literature on massive MIMO recently in terms of throughput, energy efficiency, and communication reliability \cite{BMM:15:WC, NSML:14:JSAC}. In \cite{NLM:13:TCOM}, the authors show that massive MIMO can substantially reduce the effects of noise, small-scale fading and inter-user interference by using simple linear processing, including maximum-ratio (MR) or/and zero-forcing (ZF) techniques. In particular, the transmit power of each user can be made inversely proportional to the number of antennas at the base station. Furthermore, the performance of the system can be scaled up noticeably without increasing the system complexity.

Multi-way massive MIMO relay networks, which combine massive MIMO and multi-way relaying technologies, have received research interest only recently  \cite{ALP:15:TWCOM, AP:14:PIRM, ATA:13:TCOM}. These systems are expected to reap all benefits achieved from both massive MIMO and multi-way relaying networks. Therefore, they are considered as a strong candidate to offer an enormous improvement of spectral and energy efficiency \cite{ ATA:13:TCOM}. The works of \cite{ALP:15:TWCOM, AP:14:PIRM, ATA:13:TCOM} derive the achievable rates of multi-way massive MIMO systems and show that the effects of inter-user interference and noise disappear when the number of relay antennas goes to infinity. However, these works consider ZF processing at the relay which involves a complicated matrix inversion. Furthermore, these works assume perfect CSI at the relay. In massive MIMO, channel estimates' acquisition is a critical problem, and hence, the issue of imperfect channel state information should be taken into account.

Inspired by the above discussion, in this paper we consider a far more realistic scenario, namely a multi-way massive MIMO system with MR processing and imperfect CSI. We recall that the MR processing scheme is very simple and nearly optimal when the number of relay antennas is large \cite{NSML:14:JSAC}. Furthermore, maximum-ratio processing can be implemented in a distributed manner without large backhaul requirements. A complete transmission protocol under time-division duplex (TDD) operation is proposed. We derive a corresponding expression for the spectral efficiency in closed-form.  Based on this closed-form expression, the effect of the number of relay antennas, imperfect channel estimation, and the number of users is analyzed.

\textit{Notations:}  We use bold letters for matrices and vectors. The superscripts $(\cdot)^T$, $(\cdot)^*$, and $(\cdot)^H$ denote the transpose, conjugate, and Hermitian, respectively. The symbol $\|\cdot\|$ indicates the norm of a vector. The notations $\mathbb{E}\{\cdot\}$ and $\mathbb{V}$ar$\{\cdot\}$ are the expectation and the variance operators, respectively; $\B{[X]}_{mn}$ or $x_{mn}$ denotes the $(m,n)$-th entry of matrix $\B{X}$, and $\B{I}_{K}$ is the $K\times K$ identity matrix. Moreover, $\B{[X]}_{k}$ or $\B{x}_{k}$ denotes the $k$-th column of matrix $\B{X}$. Finally, $\B{x}\sim \mathcal{CN}(0,\B{D})$ represents a circularly symmetric complex Gaussian vector $\B{x}$ with zero mean and covariance matrix $\B{D}$.

\section{System model}

\begin{figure}[t]
	\includegraphics[scale=0.17]{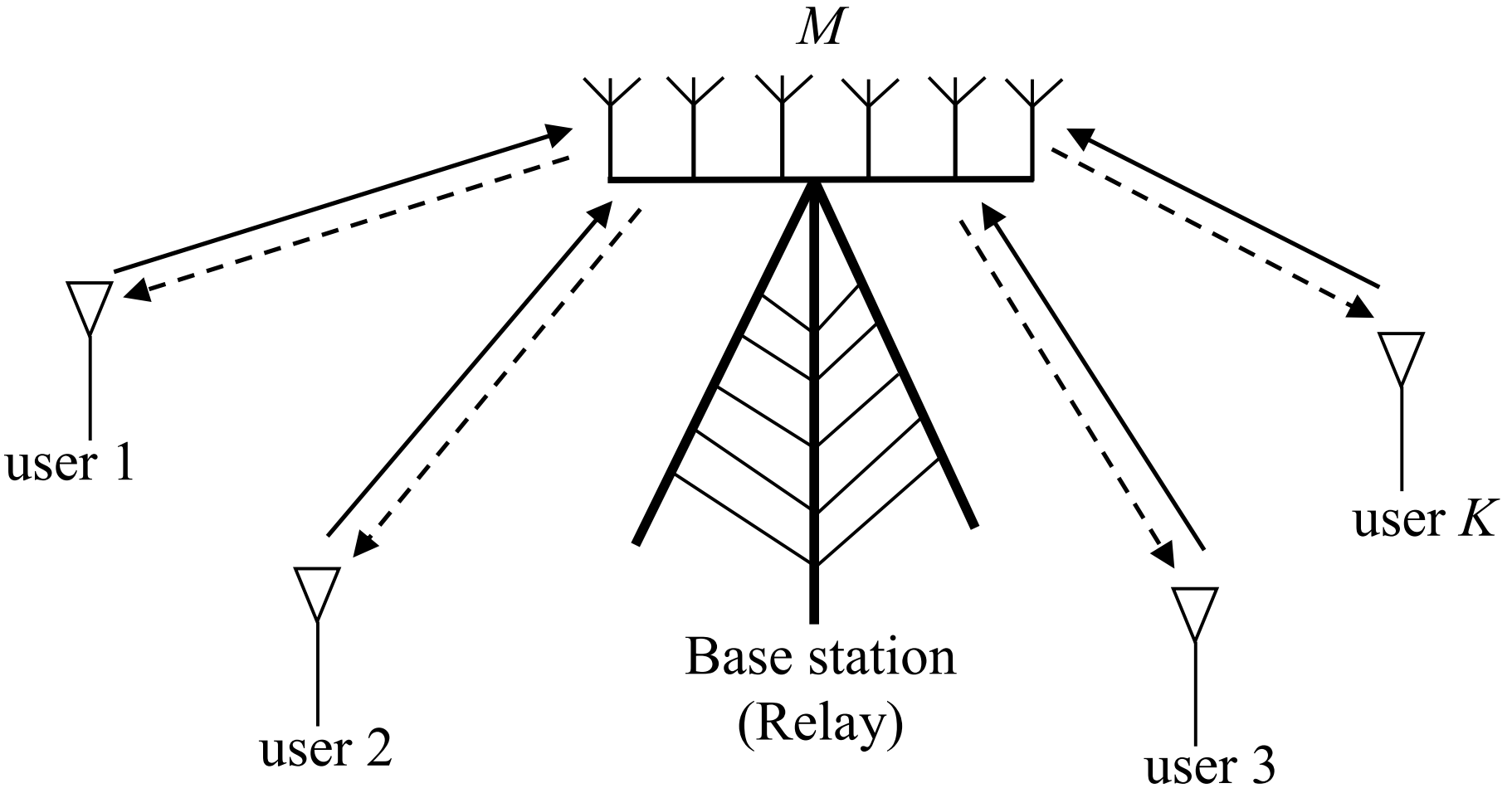}
	\caption{
		Schematic illustration of a multi-way massive MIMO system.
	}\label{fig:system_model}
\end{figure}

\begin{figure}[t]
	\includegraphics[scale=0.17]{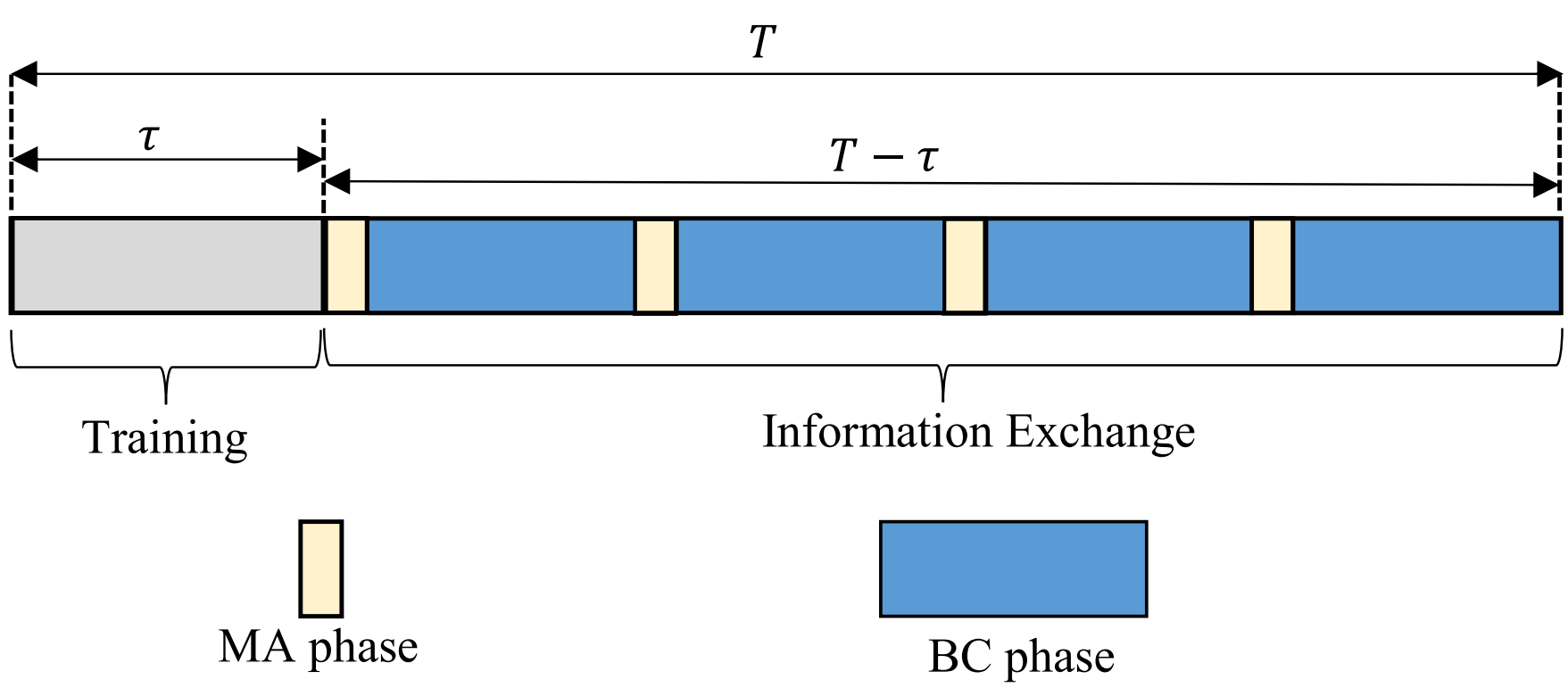}
	\caption{
		The transmission protocol  for TDD with coherence interval $T$.
	}\label{fig:tdd_model}
\end{figure}

We consider a multi-way relaying massive MIMO system which consists of one relay equipped with $M$ antennas, and $K$ single-antenna users, as shown in Figure~1. The $K$ users exchange information with each other with the help of the relay by sharing the same time-frequency resource. Each user wants to detect all $K-1$ signals transmitted from other users. We assume that all nodes operate in the half-duplex mode. Furthermore, we assume that the direct links  (user-to-user links) do not exit due to the large obstacles and/or severe shadowing. Here, we consider the scenario where both $M$ and $K$ are large, and $M$ is much larger than $K$ ($1\ll K \ll M$).

Denote by $\G$ the $M\times K$ channel matrix  between the relay and the $K$ users. In addition, $\G$ models independent small-scale fading (Rayleigh fading) and large-scale fading (geometric attenuation and log-normal shadow fading). Specifically, the $(m,k)$-th element of $\G$ is the channel coefficient between the $m$-th antenna of the relay and the $k$-th user modeled as  
  \begin{align}
  g_{mk}=h_{mk}\sqrt{\beta_{k}},
  \end{align}
 where $h_{mk}\sim \CN(0,1)$ represents the small-scale fading, and $\beta_{k}$ represents the large-scale fading.
 In matrix form, 
 \begin{align}\label{eq:G}
 \B{G}=\B{H}\B{D}^{1/2},
 \end{align}
  where $\B{H}$ is an $M\times K$ matrix, $[\B{H}]_{mk}=h_{mk}$ and $\B{D}$ is a $K\times K$ diagonal matrix, where  $[\B{D}]_{kk}=\beta_k$.

The transmission leverages TDD operation, and is divided into three phases: i) channel estimation; ii) multiple-access (MA); and iii) broadcast (BC) phases.\footnote{The schematic system diagram of information exchange for $K$ users shown in Fig. 2 implies that the system spends only one time-slot for the MA phase and $(K-1)$ times-slots for the BC phase.}

\subsection{Channel Estimation Phase}

 The relay node needs to know the channel for signal processing. To do this,  a part of coherence interval is used for channel estimation. For each coherence interval of length $T$ symbols, all users simultaneously transmit pilot sequences of length $\tau$ symbols to the relay. Let $\pmb{\phi}_k\in \mathbb{C}^{\tau\times 1}$ be the pilot sequence sent from the $k$-th user. We assume that $\pmb{\phi}_1, \pmb{\phi}_2, \dots, \pmb{\phi}_K$ are unit norm vectors and pairwisely orthogonal, i.e., $\pmb{\phi}_k^H \pmb{\phi}_{k'}=0$ for $k\neq k'$. This requires that $\tau \geq K$.

 The $M\times \tau$ received pilot matrix at relay is given by
\begin{align}
\yp
	= 
	\sum_{k=1}^K \sqrt{\tau\pp}\B{g}_k \pmb{\phi}_k^H +\B{\np}
	=
\sqrt{\tau\pp}\ \B{G}\B{\Phi}^H+\B{\np},
\end{align}
where $\B{\Phi} \triangleq [\pmb{\phi}_1, \pmb{\phi}_2, \dots, \pmb{\phi}_K]$, $\pp$ is the transmit power of each pilot symbol, $\g_k$ is the $k$-th column of $\G$, and $\np$ is the additive white Gaussian noise (AWGN) matrix with i.i.d. $\mathcal{CN}(0,1)$ components. 

At the relay, we apply the minimum mean-square-error (MMSE) technique to estimate the channel matrix $\G$ \cite{Kay:93:ECNJ}. 
The MMSE channel estimate of $\G$ is
\begin{align}
\hat{\G}=\frac{1}{\sqrt{\tau\pp}}\yp\B{\Phi}\B{\tilde{D}}=\left(\G+\frac{1}{\sqrt{\tau\pp}}\B{\tilde{N}}_\text{p}\right)\B{\tilde{D}},
\end{align}
where $\B{\tilde{D}}\triangleq\left(\frac{\B{D}^{-1}}{\tau\pp}+\B{I}_K\right)^{-1}$ and  $\mathbf{\tilde{N}}_\text{p}\triangleq\np\B{\Phi}$. From the property of $\B{\Phi}$, the elements of $\B{\tilde{N}}_\text{p}$ are i.i.d. $\mathcal{CN}(0,1)$ random variables (RVs). Let $\B{E}$ be the estimation error matrix. Then,
\begin{equation}
\G=\hat{\G}+\B{\tt{\B{E}}}.
\end{equation}
From the property of MMSE estimation, $\hat{\G}$ and $\B{E}$ are independent. Moreover, we have $\hat{\G}\sim \mathcal{CN}(0,\hat{\B{D}})$ and $\B{E}\sim \mathcal{CN}(0,\B{D}_{\text{E}})$, where $\hat{\B{D}}$ and $\B{D}_{\text{E}}$ are diagonal matrices with 
\begin{align}\label{eq:D_kk}
\left[\hat{\B{D}}\right]_{kk}=\sigma^2_k=\frac{\tau\pp\beta^2_k}{\tau\pp\beta_k+1},
\end{align}
and
\begin{align}\label{eq:D_E_kk}
\left[\B{D}_{\text{E}}\right]_{kk}=\sigma^2_{e,k}=\beta_{k}-\sigma^2_{k}.
\end{align}

\subsection{Multiple-Access Phase}

In this phase, data is transmitted to the relay in the same time-frequency resource from all users. The $M\times 1$ received vector at the relay is
\begin{align}\label{eq:yr}
\B{y}_\text{R}=\sqrt{\pu}\B{G}\B{x}+\B{n},
\end{align}
where $\B{x}=[x_{1},\dots, x_{K}]^T$, with $\mathbb{E}\left\{\left|x_{k}\right|^2\right\}=1$, is the $K\times 1$ signal vector transmitted from the $K$ users, $\B{n}$ is an $M\times 1$ AWGN vector with i.i.d. $\mathcal{CN}(0,1)$ components, and $\pu$ is the transmit power of each user. Then, the relay uses the channel estimate in the channel estimation  phase and employs the MR combining scheme as:
\begin{align}\label{eq:tilde_yr}
\tilde{\B{y}}_{\text{R}}=\hat{\B{G}}^H\yr.
\end{align}

\subsection{Broadcast Phase}

In this phase, the relay spends $K-1$ time-slots to transmit all signals to $K$ users. The relay utilizes the MR technique to broadcast a permuted version of $\tilde{\B{y}}_{\text{R}}$ at each time-slot \cite{AP:14:PIRM}. The transmit signal vector at the relay for the $t$-th $(t= 1, 2,\dots, K-1)$ time-slot can be expressed as
\begin{align}\label{eq:xr}
\B{s}_\text{R}^{(t)}&=\sqrt{\alpha^{(t)}}\hat{\B{G}}^*\B{\Pi}^{(t)}\tilde{\B{y}}_{\text{R}}\nonumber\\
&=\sqrt{\pu\alpha^{(t)}}\B{A}^{(t)}\B{x}+\sqrt{\alpha^{(t)}}\B{B}^{(t)}\B{n},
\end{align}
where $\B{\Pi}^{(t)}\in \mathbb{C}^{K\times K}$ is the permutation matrix for time slot $t$ given as \cite{ATA:13:TCOM},
\begin{equation}
\B{\Pi}^{(t)}= \begin{bmatrix}
0&  1&  0&  \cdots&  0& 0\\ 
0&  0&  1&  \cdots&  0& 0\\ 
\vdots&  \vdots&  \vdots&  \ddots&  \vdots& \vdots\\ 
0&  0&  \cdots&  \cdots&  1& 0\\ 
0&  0&  \cdots&  \cdots&  0& 1\\ 
1&  0&  \cdots&  \cdots&  0& 
0\end{bmatrix}^{t},
\end{equation}
and
	\begin{align}
	\B{A}^{(t)}&=\hat{\B{G}}^*\B{\Pi}^{(t)}\hat{\B{G}}^H\B{G},\label{A} \\
	\B{B}^{(t)}&=\hat{\B{G}}^*\B{\Pi}^{(t)}\hat{\B{G}}^H.\label{B}
	\end{align}
In \eqref{eq:xr}, $\alpha^{(t)}$ is the normalization factor, chosen to satisfy a long-term power constraint at the relay,
\begin{equation}\label{eq:pc_relay}
\mathbb{E}\left\{\left\|  \B{s}_\text{R}^{(t)} \right\|^2\right\}=\pr.
\end{equation}
From \eqref{eq:yr}, \eqref{eq:tilde_yr}, \eqref{eq:xr}, and \eqref{eq:pc_relay} the normalization factor $\alpha^{(t)}$ can be expressed as \begin{align}\label{eq:alpha_t1}
\alpha^{(t)}=\frac{\pr}{\pu\sum\limits_{k=1}^{K}{\mathbb{E}\left\{\left\|\left[\B{A}^{(t)}\right]_{k}\right\|^2\right\}}+\sum\limits_{m=1}^{M}{\mathbb{E}\left\{\left\|\left[\B{B}^{(t)}\right]_{m}\right\|^2\right\}}}.
\end{align}

Then, the $K\times 1$ received  signal vector at the $K$ users in the $t$-th time slot can be described as follows:
\begin{equation}\label{eq:yu}
\yu^{(t)}=\G^T\B{s}_\text{R}^{(t)}+\B{w}^{(t)},
\end{equation}
and substituting \eqref{eq:xr} into \eqref{eq:yu}, we obtain
\begin{align}\label{yu_AB}
\yu^{(t)}=\sqrt{\alpha^{(t)}\pu}\B{G}^T\B{A}^{(t)}\B{x}+\sqrt{\alpha^{(t)}}\B{G}^T\B{B}^{(t)}\B{n}+\B{w}^{(t)}.
\end{align}

\section{Spectral Efficiency Analysis}

In this section, we analyse the spectral efficiency of the system. More specifically, we derive a closed-form expression for the spectral efficiency. Without loss of generality, we analyze the performance of the system for the first time-slot. The performance analysis for other time-slots follows the same methodology. Note that, hereafter, if $k=K$, then we set $k+1=1$ and $k+2=2$; if $k=1$, then we set $k-1=K$; and if $k=2$, we set $k-2=K$.

In the first time-slot, the $k$-th user wants to detect the signal $x_{k+1}$ transmitted from the $(k+1)$-th user. From \eqref{yu_AB}, the received signal in the first time slot for the $k$-th user is described by
\begin{align}
y^{(1)}_{\text{u},k}&=\sqrt{\alpha^{(1)}\pu}\ \B{g}^T_k\B{a}^{(1)}_{k+1}x_{k+1} +\sqrt{\alpha^{(1)}\pu}\!\!\!\!\sum_{\overset{i=1}{i\neq(k+1)}}^{K}\!\!\!\!\B{g}_k^T\B{a}_i^{(1)}x_i \nonumber\\
&+\sqrt{\alpha^{(1)}}\sum_{m=1}^{M}\B{g}^T_k\B{b}_m^{(1)}{n}_m+ w_k^{(1)}\label{eq:y1}\\
&=\underset{\text {desired signal}}{\underbrace{\sqrt{\alpha^{(1)}\pu} \mathbb{E}\left\{\B{g}^T_k\B{a}^{(1)}_{k+1}\right\}x_{k+1}}}+\underset{\text {effective noise}}{\underbrace{\tilde{N}_{k}^{(1)}}},\label{eq:y11}
\end{align}
where $\tilde{N}_{k}^{(1)}$ is considered as the effective noise and given by

\begin{align}
&\tilde{N}_{k}^{(1)}= \underset{\text {beamforming uncertainty}}{\underbrace{\sqrt{\alpha^{(1)}\pu}\ \left(\B{g}^T_k\B{a}_{k+1}^{(1)}- \mathbb{E}\left\{\B{g}^T_k\B{a}^{(1)}_{k+1}\right\}\right)x_{k+1}}}\nonumber\\
&+\underset{\text {inter-user interference}}{\underbrace{\sqrt{\pu\ \alpha^{(1)}}\!\!\!\!\sum_{\overset{i=1}{i\neq(k+1)}}^{K}\!\!\!\!\g_k^T\B{a}_i^{(1)}x_i}} + \underset{\text {noise}}{\underbrace{\sqrt{\alpha^{(t)}}\sum_{m=1}^{M}\g_k^T\B{b}_m^{(1)}n_{m} +w_k^{(1)}}}.\label{eq:y12}
\end{align}

\newcounter{MYtempeqncn}
\begin{figure*}
	\normalsize
	\setcounter{MYtempeqncn}{\value{equation}}
	\setcounter{equation}{23}
	\begin{align}\label{eq:alpha_t2}
	\alpha^{(1)}=\frac{\pr}{M^3\pu\sum\limits_{k'=1}^{K}\sigma^2_{k'-1}\sigma^4_{k'}+M^2\left(\sum\limits_{k'=1}^{K}\sigma^2_{k'}\sigma^2_{k'+1}\right)\left(\pu\sum\limits_{k'=1}^{K}\beta_{k'}+1\right)+M\pu\sum\limits_{k'=1}^{K}\sigma^4_{k'}\sigma^2_{k'+1}},
	\end{align}
	\setcounter{equation}{\value{MYtempeqncn}}
	\hrulefill
	\vspace*{4pt}
\end{figure*}

From \eqref{eq:y11}, it can be clearly seen that the ``desired signal" term  is uncorrelated with the ``effective noise" term. Thus, the signal-to-interference-plus-noise ratio (SINR) in the first time-slot for the $k$-th user is given by
\begin{align}\label{eq:SINR_1}
\gamma_{k}^{(1)}=\frac{\alpha^{(1)}\pu\left|\mathbb{E}\left\{\g_k^T\B{a}_{k+1}^{(1)}\right\}\right|^2}{\alpha^{(1)}\pu\mathbb{V}\text{ar}\left(\g_k^T\B{a}_{k+1}^{(1)}\right)+{\tt{IU}}_k+{\tt{AN}}_k+1},
\end{align}
where 
\begin{align}
{\tt{IU}}_k&=\alpha^{(1)}\pu\!\!\!\!\sum_{\overset{i=1}{i\neq(k+1)}}^{K}\!\!\!\!\mathbb{E}\left\{\left|\B{g}_k^T\B{a}_i^{(1)}\right|^2\right\}\label{eq:IS_k},\\
{\tt{AN}}_k&=\alpha^{(1)}\sum_{m=1}^{M}\mathbb{E}\left\{\left|\B{g}_k^T\B{b}_m^{(1)}\right|^2 \right\}.\label{eq:IB_k}
\end{align}
In \eqref{eq:SINR_1}, $\alpha^{(1)}$ is given by \eqref{eq:alpha_t1}, and it can be represented in closed-form via \eqref{eq:alpha_t2} shown at the top of the next page. The detailed derivation of \eqref{eq:alpha_t2} is shown in Appendix~\ref{app:alpha}. Then, the spectral efficiency of the $k$-th user in bit/s/Hz is given by
\setcounter{equation}{24}
\begin{align}\label{eq:SE_k}
{\tt{SE}}^{(1)}_{k}
&=\left(\frac{T-\tau}{T}\right)\left(\frac{K-1}{K}\right)\text{log}_{2}\left(1+\gamma_{k}^{(1)}\right).
\end{align}
The pre-log factor in \eqref{eq:SE_k} includes: i) $\frac{T-\tau}{T}$ which comes from the fact that during a coherence interval of $T$ symbols, we spend $\tau$ symbols for the training; and ii)  $\frac{K-1}{K}$ accounts for the fact that we spend $K$ time slots to send $K-1$ signals to a given user. We next provide a closed-form expression for the spectral efficiency given in \eqref{eq:SE_k}.

\begin{theorem}\label{theo:1}
The closed-form expression for the spectral efficiency \eqref{eq:SE_k} is 
\begin{align}\label{eq:SINR_2}
&{\tt{SE}}_{k}^{(1)}=\left(\frac{T-\tau}{T}\right)\left(\frac{K-1}{K}\right)\nonumber\\
&\times\log_{2}\left(1+\frac{\alpha^{(1)}\pu \ M^4\sigma^4_{k}\sigma^4_{k+1}}{\alpha^{(1)}\pu\mathbb{V}\text{ar}\left\{\g_k^T\B{a}_{k+1}^{(1)}\right\}+{\tt{IU}}_{k}+{\tt{AN}}_{k}+1}\right),
\end{align}
where \begin{align}
&\mathbb{V}\text{ar}\left\{\g_k^T\B{a}_{k+1}^{(1)}\right\}=M^3\ai_{k,k+1} + M^2\bi_{k,k+1} + M\ci_{k,k+1},\label{eq:var2}\\
&{\tt{IU}}_k
=
\alpha^{(1)}\pu\sum_{\overset{i=1}{i\neq(k+1)}}^{K}
\left(M^3\ai_{k,i} + M^2\bi_{k,i} + M\ci_{k,i}\right) \nonumber\\
&+ \alpha^{(1)}\pu M^2\sigma_{k-1}^4\sigma_k^4 + \alpha^{(1)}\pu M\Big[2\sigma_{k}^6\sigma_{k+1}^2 + 2\sigma_{k}^6\sigma_{k-1}^2\nonumber\\
&+\left(2\sigma_{k}^4 + \beta_k^2 - 2\beta_k\sigma_k^2\right)\sum_{k'=1}^K 2\sigma_{k'}^2\sigma_{k'+1}^2\Big]
,\label{eq:is}\\
&{\tt{AN}}_k=\alpha^{(1)}\left(M^3\sigma^4_{k}\sigma^2_{k+1}+M^2\beta_{k}\sum_{k'=1}^{K}\sigma^2_{k'}\sigma^2_{k'+1}\right),\label{eq:ib1}
\end{align} 
where
\begin{align}\label{eq:ai}
\ai_{k,i} 
&=
\sigma_k^4\sigma_{k+1}^2\beta_i +  \sigma_i^4\sigma_{i-1}^2\beta_k,\\
\bi_{k,i} 
&=
\beta_k\beta_i \sum_{k'=1}^K\sigma_{k'}^2\sigma_{k'+1}^2,\\
\ci_{k,i} 
&=
\sigma_k^4\sigma_{k-1}^2\beta_i +  \sigma_i^4\sigma_{i+1}^2\beta_k.	
\end{align}
\begin{proof}
	See Appendix~\ref{proofpro1}.
\end{proof}

\end{theorem}

Result \eqref{eq:SINR_2} yields some important remarks:
\begin{itemize}
	\item As $M\to\infty$, the numerator of the SINR of \eqref{eq:SINR_2} scales as $M$ while the denominator converges to a constant, and hence, the spectral efficiency grows without bound.
	\item If the transmit power of each user is scaled with $1/M$, i.e. $\pu=E_{\text{u}}/{M}$, where $E_\text{u}$ is fixed, then as $M\to\infty$, the spectral efficiency converges to
	\begin{align}\label{eq:se_Eu/M}
	{\tt{SE}}_{k}^{(1)}\to \left(\frac{T-\tau}{T}\right)\left(\frac{K-1}{K}\right)\log_2\left(1+E_\text{u}\sigma^2_{k+1}\right).
	\end{align}
	The result in \eqref{eq:se_Eu/M} implies that by utilizing very large number of antennas $M$ at the relay, we can cut down the transmitted power of each user proportionally to $1/M$ without reducing the system performance.
	
	\item If the transmit power at the relay is scaled with $1/M$, i.e., $\pr=E_{\text{r}}/{M}$, where $E_\text{r}$ is fixed, then as $M\to\infty$, we obtain
	\begin{align}\label{eq:se_Er/M}
	{\tt{SE}}_{k}^{(1)}\to \left(\frac{T-\tau}{T}\right)\left(\frac{K-1}{K}\right)\log_2\left(1+\frac{E_\text{r}\sigma^4_{k}\sigma^4_{k+1}}{\sum\limits_{k'=1}^{K}\sigma^2_{k'-1}\sigma^4_{k'}}\right),
	\end{align}
	which indicates that we can reduce the transmitted power at the relay proportionally to $1/M$, when $M$ is large.
	
	\item If the transmit power of the relay and each user are scaled with $1/M$, i.e. $\pr=E_{\text{r}}/{M}$, and $\pu=E_{\text{u}}/{M}$, where $E_\text{r}$ and $E_\text{u}$ are fixed, then as $M \to \infty$, the spectral efficiency converges to
	\begin{align}\label{eq:se_E_u_Er/M}
	{\tt{SE}}_{k}^{(1)}\to \left(\frac{T-\tau}{T}\right)\left(\frac{K-1}{K}\right)\log_2\left(1+\frac{\xi E_\text{u}\sigma^4_{k}\sigma^4_{k+1}}{\xi\sigma^4_{k}\sigma^2_{k+1}+1}\right),
	\end{align}
	where 
	\begin{align}
	\xi=\frac{E_\text{r}}{\sum\limits_{k'=1}^{K}\left(E_\text{u}\sigma^2_{k'-1}\sigma^4_{k'}+\sigma^2_{k'}\sigma^2_{k'+1}\right)}.
	\end{align}
We can see that, when $M$ is large, the transmit power at the relay and each user can be cut down proportionally $1/M$, while maintaining a non-zero spectral efficiency.
\end{itemize}

\section{Numerical Results}

In this section, numerical results are presented to verify our analysis. For all examples, we choose $T=200$, and define ${\tt SNR}= P_\text{u}$ (expressed in dB). We examine the sum spectral efficiency as follows:
\begin{align}\label{eq:sum_spect_a}
{\tt{SE}}_{\text{sum}}=\sum_{k=1}^{K}{\tt{SE}}^{(1)}_{k}.
\end{align}
\subsection{Scenario I}
\begin{figure}[t]
	\includegraphics[scale=0.26]{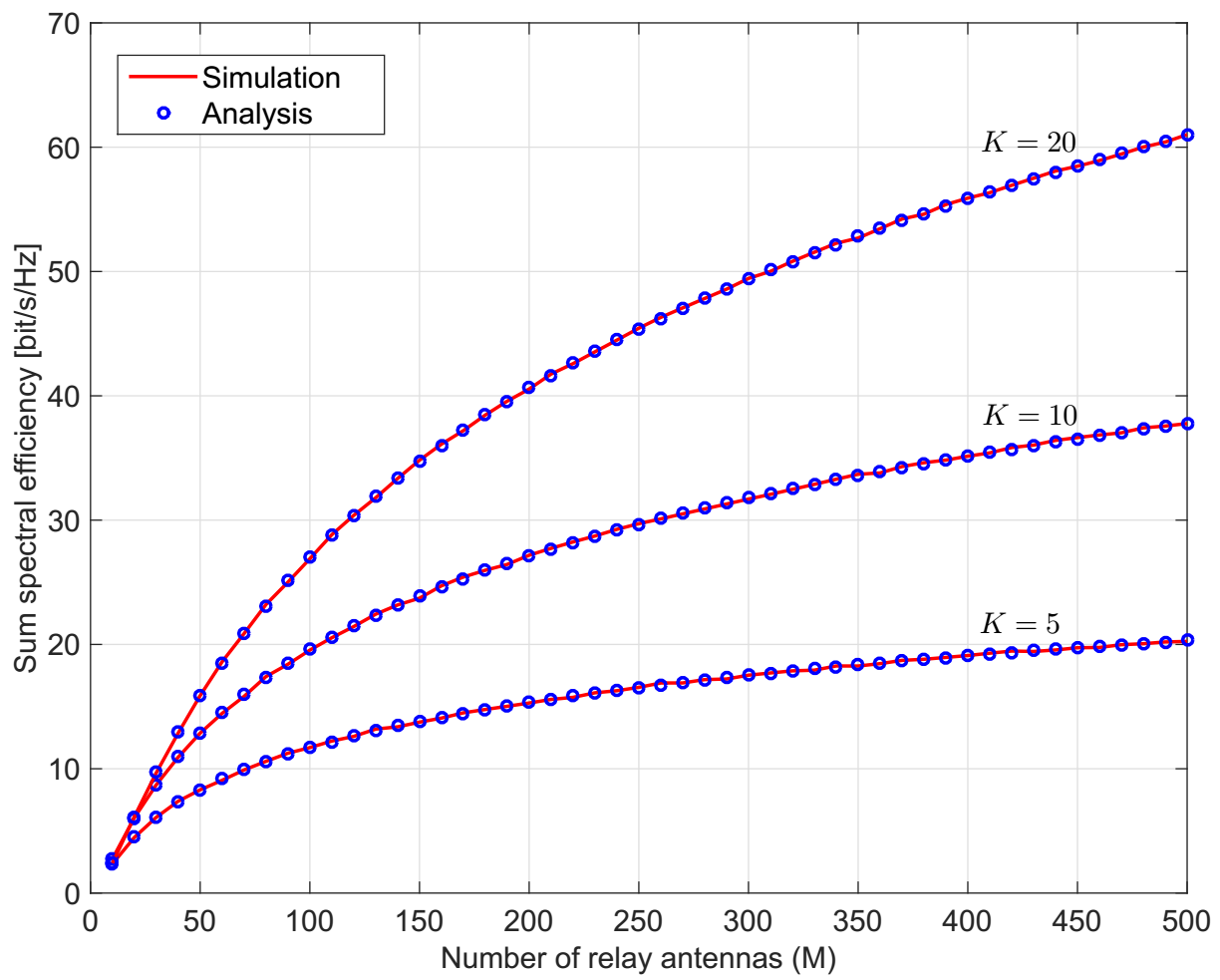}
	\caption{
		Sum spectral efficiency versus the number of relay antennas. We choose $T=200$, $\tau=K$, $\pp=\pu=0$ dB, $\pr=10$ dB, and $\beta_{k}=1$.
	}\label{fig:spectral_efficiency10}
\end{figure}

We consider a simple case where the large-scale fading $\beta_{k}=1$ for all $k$. Figure~\ref{fig:spectral_efficiency10} shows the sum spectral efficiency versus $M$ for different $K$. The solid and circle lines present the Monte-Carlo simulations using  \eqref{eq:SE_k} and analytical results using \eqref{eq:SINR_2}, respectively. From the figure, we can see that the analytical results match the simulation results which validates the correctness of our closed-form expression. As expected, when $M$ increases the sum spectral efficiency increases. Furthermore, for low numbers of antennas, for example $M=20$, the gaps of the sum spectral efficiency between the three cases $(K=5, 10$, and $20)$ are very small. However, these gaps grow considerably when the number of antennas is large, i.e., at $M=500$, the sum spectral efficiency with $K=10$ is nearly double compared to the one with $K=5$. This is due to the fact that when $M$ increases, inter-user interference and noise effects are canceled out. 

\begin{figure}[t]
	\includegraphics[scale=0.26]{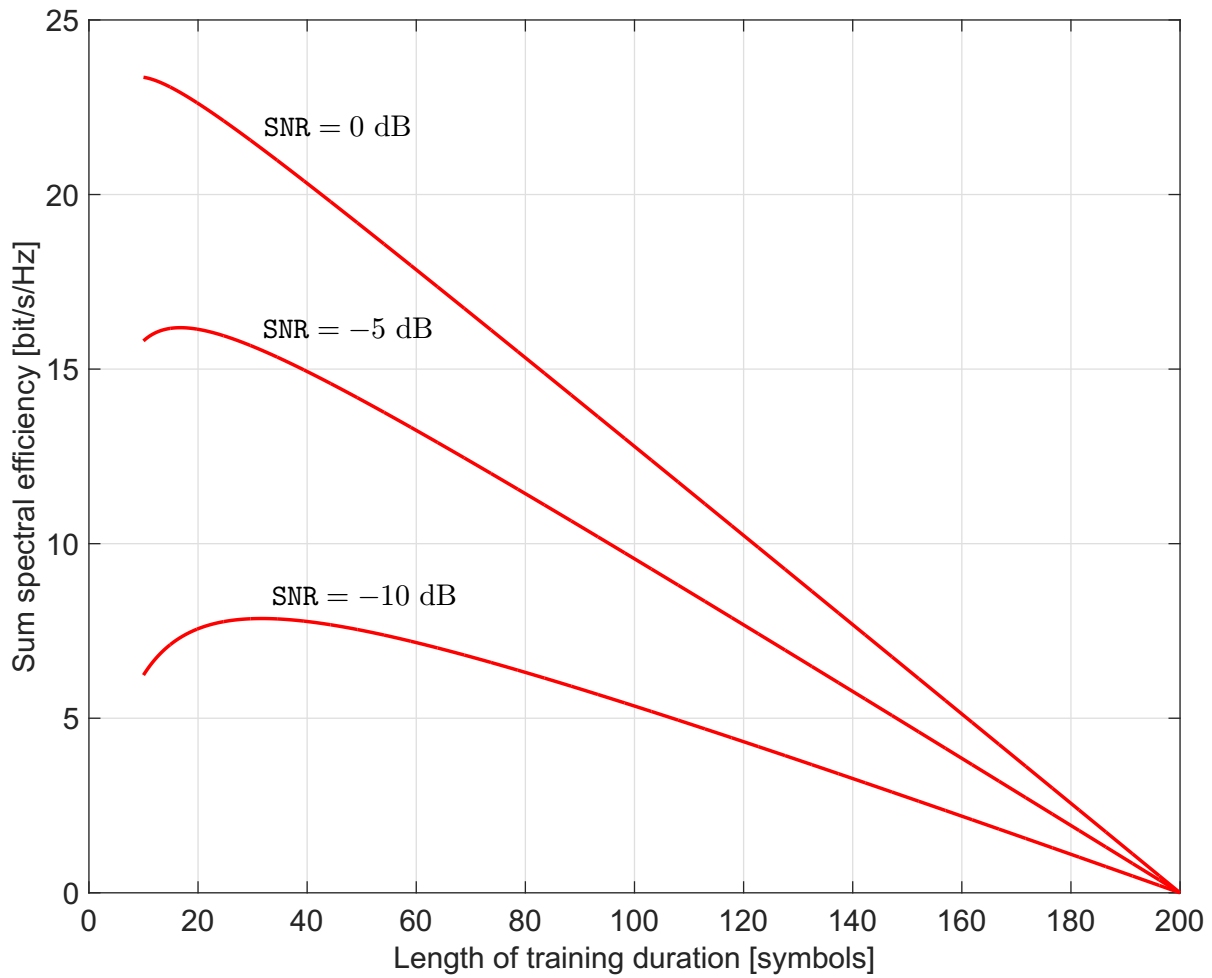}
	\caption{
		Sum spectral efficiency versus the length of training duration. We choose, $M=200$, $K=10$, $T=200$, $\pr=\pp=10$ dB, and $\beta_{k}=1$.
	}\label{fig:pilot_sequences}
\end{figure}

\begin{figure}[t]
	\includegraphics[scale=0.26]{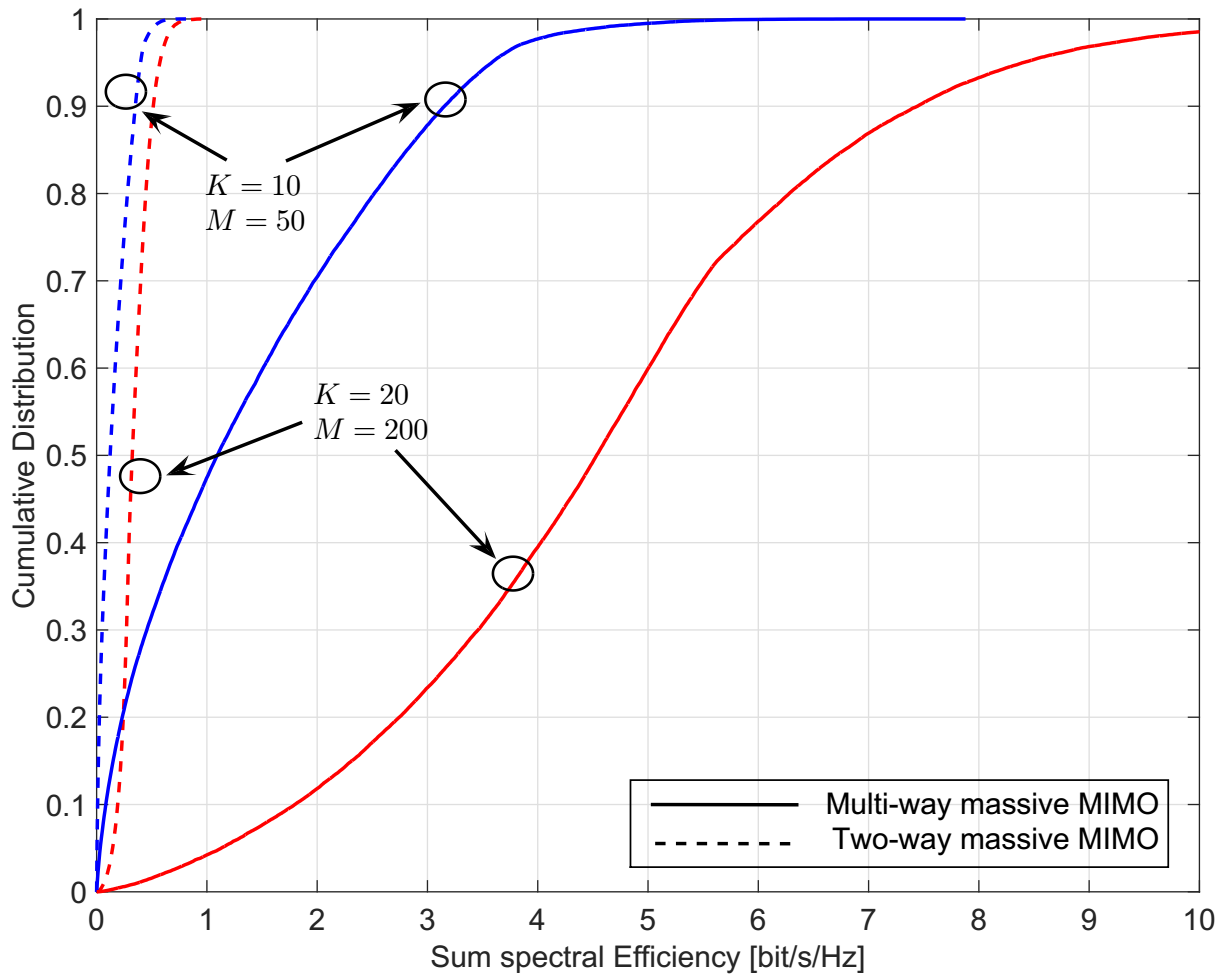}
	\caption{
		Cumulative distribution of the sum spectral efficiency. We choose $D_d=1000$ m, $T=200$, $\tau=K$, $\pu=\pr=\pp=0$ dB.
	}\label{fig:cumulative_distribution_different_K_M}
\end{figure}

Next, we investigate the effect of the length of training duration $\tau$ on the system performance, as shown in Figure~\ref{fig:pilot_sequences}. The figure shows the sum spectral efficiency versus the length of training pilot duration for different ${\tt SNR}$. When $\tau$ increases, the channel estimation is more accurate, but the channel estimation overhead increases (which reduces the pre-log factor of \eqref{eq:SINR_2}). At short $\tau$ (e.g. $10\leqslant\tau\leqslant 20$), the length of information exchange duration is relatively large compared to the length of training duration, and hence, the system performance improves with $\tau$. By contrast, when $\tau$ is large, channel estimation overhead dominates the system performance, and therefore, the sum spectral efficiency reduces when $\tau$ increases. There exist optimal values of $\tau$ so that the sum spectral efficiencies are maximal. These values increase when $\tt SNR$ decreases. The reason is that when the $\tt SNR$ is small, we need to use more pilot symbols to increase the channel estimation accuracy. This observation is in line with \cite{ngo2014massive}.

\subsection{Scenario II}

We now examine a more practical scenario that takes into account a realistic large-scale fading model. More precisely, the large-scale fading, is modeled by path loss and shadowing, given by \cite{NSML:14:JSAC}
\begin{align}
\beta_{k}=\frac{z_{k}}{1+\left(\frac{d_{k}}{d_{0}}\right)^\nu},
\end{align}
where $z_{k}$ is the log-normal random variable with standard deviations of $\sigma_{z}$ dB, $\nu$ represents the path loss exponent, $d_{k}$ is the distance between user $k$ and the relay, and $d_{0}$ denotes a reference distance. Here, we assume that users are located uniformly at random inside a disk with a diameter $D_{d}$. 

For our comparison, we also consider two-way  massive MIMO relaying systems where the $K$ users are grouped into $K(K-1)/2$ pairs, and each is assigned different time-slots and uses the two-way relaying scheme to exchange data.\footnote{Alternatively, one could consider a multi-pair two-way relaying protocol as in \cite{KZMEZ:16:CIT}, where $K$ sources exchange data with $K$ destinations over two orthogonal time-slots (i.e., the number of users is always an even number). For this case, the pre-log penalty is $1/2$. Compared with our multi-way relaying protocol, the multi-pair two-way relaying protocol has a smaller pre-log factor (when $K>2$), but suffers a similar interference effect since multiple users transmit their data in the same time-frequency resource.} Since each pair requires two time-slots for information exchange, we need, in total, $K(K-1)$ time-slots to exchange all information among the $K$ users. As a result, the spectral efficiency of the two-way massive MIMO relaying system is given by
\begin{align}\label{eq:two-way}
		{\tt{SE}}_{\text{two-way},k}^{(1)}
		&=\left(\frac{T-\tau}{T}\right)\left(\frac{K-1}{K(K-1)}\right)\log_2\left(1+{\tt SINR}_k\right)\nonumber\\
		&= \left(\frac{T-\tau}{T}\right)\left(\frac{1}{K}\right)\log_2\left(1+{\tt SINR}_k\right),
\end{align}
where ${\tt SINR}_k$ corresponds to the SINR in \eqref{eq:SINR_2} for $K=2$.

In this example, we choose  $D_{d}=1000$ m, $\sigma_{z}=8$ dB, $\nu=4$, $d_{0}=200$ m, $\pu=\pr=\pp=0$ dB. Figure~\ref{fig:cumulative_distribution_different_K_M} demonstrates the cumulative distribution of the sum spectral efficiencies for two cases: $(K=20, M=200)$ and $(K=10, M=50)$. Compared to the two-way massive MIMO system, the multi-way massive MIMO system reduces the pre-log penalty (from $\frac{1}{K}$ to $\frac{K-1}{K}$), however, it exhibits more interference since many users simultaneously transmit data in the same frequency band. When $M$ is large, the interference is small. Therefore, the multi-way massive MIMO system outperforms the two-way massive MIMO system, especially at large $M$.

\section{Conclusion}

We studied multi-way massive MIMO relay networks with MR processing and imperfect CSI. We derived a closed-form expression for the spectral efficiency. Our work showed that, by using a large antenna array at the relay, many users can simultaneously  exchange their information in the same frequency band without any performance degradation for each user. As a result, multi-way massive MIMO offers much higher sum spectral efficiency  than conventional multi-way MIMO or two-way massive MIMO systems do. Furthermore, as the number of relay antennas grows large, we can reduce the transmitted power at the relay and/or each user proportionally to $1/M$, while maintaining a given quality-of-service.

\section{Appendices}

\subsection{Derivation of \eqref{eq:alpha_t2}} \label{app:alpha}

From \eqref{A} and \eqref{B}, the normalization factor $\alpha^{(1)}$ in \eqref{eq:alpha_t1} can be rewritten as
\begin{align}\label{eq:alpha_t22}
\alpha^{(1)}
	=
	\frac{\pr}{\pu\sum\limits_{k=1}^{K}{\tt{Q_1}}_{k}
		+ \pu\sum\limits_{k=1}^{K}{\tt{Q_2}}_{k}
		+\sum\limits_{m=1}^{M}{\tt{Q_3}}_{m}},
\end{align}
where
\begin{align}
{\tt{Q_1}}_{k}&=\mathbb{E}\left\{\left\|\left[\Gh^*\B{\Pi}^{(1)}\Gh^H\Gh\right]_{k}\right\|^2\right\},\\
{\tt{Q_2}}_{k}&=\mathbb{E}\left\{\left\|\left[\Gh^*\B{\Pi}^{(1)}\Gh^H\B{E}\right]_{k}\right\|^2\right\},\\
{\tt{Q_3}}_{m}&=\mathbb{E}\left\{\left\|\left[\Gh^*\B{\Pi}^{(1)}\Gh^H\right]_{m}\right\|^2\right\}.
\end{align}
First, we compute ${\tt{Q_1}}_{k}$.
We have,
\begin{align}\label{eq:q1st}
{\tt{Q_1}}_{k}
&=\mathbb{E}\left\{\left\|\left[\Gh^*\B{\Pi}^{(1)}\Gh^H\Gh\right]_{k}\right\|^2\right\}\nonumber\\
&=\sum_{k'=1}^{K}\mathbb{E}\left\{\left\|\gh^*_{k'}\gh^H_{k'+1}\gh_{k}\right\|^2\right\}\nonumber\\
&=\mathbb{E}\left\{\left\|\frac{\gh^H_{k+1}\gh_{k}}{\|\gh_{k}\|}\|\gh_{k}\|\gh_{k}^*\right\|^2\right\}+\mathbb{E}\left\{\left\|\gh^*_{k-1}\right\|^2\|\gh_{k}\|^4\right\}\nonumber\\
&+\!\!\!\!\!\!\!\!\sum\limits_{\overset{k'=1}{k'\neq(k,k-1)}}^{K}\!\!\!\!\!\!\!\!\left\{\left\|\gh^*_{k'}\gh^H_{k'+1}\gh_{k}\right\|^2\right\}\nonumber\\
&=\sigma^2_{k+1}\mathbb{E}\left\{\|\gh_{k}\|^4\right\}+M\sigma^2_{k-1}\mathbb{E}\left\{\|\gh_{k}\|^4\right\}\nonumber\\
&+\!\!\!\!\!\!\!\!\sum\limits_{\overset{k'=1}{k'\neq(k,k-1)}}^{K}\!\!\!\!\!\!\!\!\left\{\left\|\gh^*_{k'}\gh^H_{k'+1}\gh_{k}\right\|^2\right\},
\end{align}
where in the last equality we have used the fact that $\frac{\gh^H_{k+1}\gh_{k}}{\|\gh_{k}\|} \sim \mathcal{CN}(0,\sigma_{k+1}^2)$ is independent of $\gh_{k}$ \cite{NLM:13:TCOM}.
By using Lemma 2.9 in \cite{AMT:04:CIT}, \eqref{eq:q1st} becomes
\begin{align}\label{eq:q1}
&{\tt{Q_1}}_{k}=M(M+1)\sigma^4_{k}\sigma^2_{k+1}+M^2(M+1)\sigma^2_{k-1}\sigma^4_{k}\nonumber\\
&+M^2\sigma^2_{k}\!\!\!\!\!\!\!\!\sum\limits_{\overset{k'=1}{k'\neq(k,k-1)}}^{K}\!\!\!\!\!\!\!\!\sigma^2_{k'}\sigma^2_{k'+1}\nonumber\\
&=M^3\sigma^2_{k-1}\sigma^4_{k}+M\sigma^4_{k}\sigma^2_{k+1}+M^2\sigma^2_{k}\sum_{k'=1}^{K}\sigma^2_{k'}\sigma^2_{k'+1}.
\end{align}
Similarly, we obtain
\begin{align}
&{\tt{Q_2}}_{k}=M^2(\beta_{k}-\sigma^2_{k})\sum_{k'=1}^{K}\sigma^2_{k'}\sigma^2_{k'+1}\label{eq:q2},
\end{align}
and
\begin{align}
{\tt{Q_3}}_{m}&=M\sum_{k'=1}^{K}\sigma^2_{k'}\sigma^2_{k'+1}.\label{eq:q3}
\end{align}
Substituting \eqref{eq:q1}, \eqref{eq:q2} and \eqref{eq:q3} into \eqref{eq:alpha_t1}, we obtain \eqref{eq:alpha_t2}.

\subsection{Proof of Theorem~\ref{theo:1}}\label{proofpro1}

$1)$ Compute $\mathbb{E}\left\{\g_k^T\B{a}_{k+1}^{(1)}\right\}$: 
We have,
\begin{align}
&\mathbb{E}\left\{\g_k^T\B{a}_{k+1}^{(1)}\right\}\nonumber\\
&=\mathbb{E}\left\{\left(\gh^T_k+\e^T_{k}\right)\left(\left[\hat{\G}^*\B{\Pi}^{(1)}\hat{\G}^H(\Gh+\B{E})\right]_{k+1}\right)\right\}.
\end{align}

Since $\Gh$ and $\B{E}$ are independent, we obtain

\begin{align}
&\mathbb{E}\left\{\g_k^T\B{a}_{k+1}^{(1)}\right\}=\mathbb{E}\left\{\g^T_k\left[\hat{\G}^*\B{\Pi}^{(1)}\Gh^H\Gh\right]_{k+1}\right\}\nonumber\\
&=\mathbb{E}\left\{\|\gh_{k}\|^2\|\gh_{k+1}\|^2\right\}+ \sum\limits_{\overset{k'=1}{k'\neq k}}^{K}\mathbb{E}\left\{\gh^T_{k}\gh^*_{k'}\gh^H_{k'+1}\gh_{k+1}\right\}\nonumber\\
&= M^2\sigma^2_{k}\sigma^2_{k+1}\label{eq:TS2}.
\end{align}

$2)$ Compute $\mathbb{V}\text{ar}\left(\g_k^T\B{a}_{k+1}^{(1)}\right)$:
From  \eqref{eq:TS2}, the variance of $\g_k^T\B{a}_{k+1}^{(1)}$ is given by
\begin{align}
&\mathbb{V}\text{ar}\left(\g_k^T\B{a}_{k+1}^{(1)}\right)\nonumber
=\mathbb{E}\left\{\left|\g_k^T\B{a}_{k+1}^{(1)}\right|^2\right\}-\left|\mathbb{E}\left\{\g_k^T\B{a}_{k+1}^{(1)}\right\}\right|^2\\
&=\mathbb{E}\left\{\left|\g_k^T\B{a}_{k+1}^{(1)}\right|^2\right\}-M^4\sigma_k^4\sigma_{k+1}^4\nonumber\\
&=
\mathbb{E}\left\{\left|\left(\hat{\g}_k^T+\e_k^T\right)\Gh^*\B{\Pi}^{(1)}\hat{\G}^H\left(\hat{\g}_{k+1}+\e_{k+1}\right)\right|^2\right\}\nonumber\\
&-M^4\sigma_k^4\sigma_{k+1}^4.\label{eq:var}
\end{align}
Since $\Gh$ and $\B{E}$ are independent, \eqref{eq:var} can be rewritten as
\begin{align}
\mathbb{V}\text{ar}\left(\g_k^T\B{a}_{k+1}^{(1)}\right)
=
{\tt{T_1}} + {\tt{T_2}} + {\tt{T_3}} + {\tt{T_4}}-M^4\sigma_k^4\sigma_{k+1}^4,\label{eq:var1}
\end{align}
where 
\begin{align}
{\tt{T_1}}
	&=
\mathbb{E}\left\{\left|\gh_k^T\Gh^*\B{\Pi}^{(1)}\Gh^H\gh_{k+1}\right|^2\right\},\label{eq:T1}\\
{\tt{T_2}}
	&=
\mathbb{E}\left\{\left|\gh_k^T\Gh^*\B{\Pi}^{(1)}\Gh^H\e_{k+1}\right|^2\right\},\\
{\tt{T_3}}
	&=
\mathbb{E}\left\{\left|\e_k^T\Gh^*\B{\Pi}^{(1)}\Gh^H\gh_{k+1}\right|^2\right\},\\
{\tt{T_4}}
	&=
\mathbb{E}\left\{\left|\e_k^T\Gh^*\B{\Pi}^{(1)}\Gh^H\e_{k+1}\right|^2\right\}.
\end{align}
To compute ${\tt{T_1}}$, we rewrite \eqref{eq:T1} as
\begin{align}\label{eq:t11}
&{\tt{T_1}}=\sum_{k'=1}^{K}\mathbb{E}\left\{\left|\gh_k^T\gh^*_{k'}\gh^H_{k'+1}\gh_{k+1}\right|^2\right\}\nonumber\\
&=\mathbb{E}\left\{\left\|\gh_k\right\|^4\right\}\mathbb{E}\left\{\left\|\gh_{k+1}\right\|^4\right\}+\mathbb{E}\left\{\left|\gh_k^T\gh^*_{k+1}\gh^H_{k+2}\gh_{k+1}\right|^2\right\}\nonumber\\
&+\mathbb{E}\left\{\left|\gh_k^T\gh^*_{k-1}\gh^H_{k}\gh_{k+1}\right|^2\right\}+\!\!\!\!\!\!\!\!\!\!\!\!\!\!\sum\limits_{\overset{k'=1}{k'\neq(k,k-1,k+1)}}^{K}\!\!\!\!\!\!\!\!\!\!\!\mathbb{E}\left\{\left|\gh_k^T\gh^*_{k'}\gh^H_{k'+1}\gh_{k+1}\right|^2\right\}.
\end{align}
Again, by using Lemma 2.9 in \cite{AMT:04:CIT}, we get
\begin{align}\label{eq:t1}
{\tt{T_1}}&=M^2(M+1)^2\sigma_k^4\sigma_{k+1}^4+M(M+1)\sigma_k^2\sigma_{k+1}^4\sigma_{k+2}^2\nonumber\\
&+M(M+1)\sigma^2_{k-1}\sigma^2_{k+1}\sigma_k^4+M^2\!\!\!\!\!\!\!\!\!\!\!\!\!\sum\limits_{\overset{k'=1}{k'\neq(k,k-1,k+1)}}^{K}\!\!\!\!\!\!\!\!\!\!\!\sigma_k^2\sigma^2_{k'}\sigma^2_{k'+1}\sigma^2_{k+1}\nonumber\\
&=M^3(M+2)\sigma_k^4\sigma^4_{k+1}+M\sigma^2_{k-1}\sigma^4_k\sigma^2_{k+1}\nonumber\\
&+M\sigma^2_{k}\sigma^4_{k+1}\sigma^2_{k+2}+M^2\sigma^2_{k}\sigma^2_{k+1}\sum_{k'=1}^{K}\sigma^2_{k'}\sigma^2_{k'+1}.
\end{align}
Similarly, we obtain
\begin{align}
{\tt{T_2}}&=M^3\sigma_{e,k+1}^2\sigma^4_{k}\sigma^2_{k+1} + M\sigma^2_{e,k+1}\sigma^2_{k-1}\sigma^4_{k}\label{eq:t2}\nonumber\\
&+M^2\sigma^2_{k}\sigma^2_{e,k+1}\sum_{k'=1}^{K}\sigma^2_{k'}\sigma^2_{k'+1},
\end{align}

~

\begin{align}
{\tt{T_3}}&=M^3\sigma_{e,k}^2\sigma^2_{k}\sigma^4_{k+1} + M\sigma^2_{e,k}\sigma^4_{k+1}\sigma^2_{k+2}\label{eq:t3}\nonumber\\
&+M^2\sigma^2_{e,k}\sigma^2_{k+1}\sum_{k'=1}^{K}\sigma^2_{k'}\sigma^2_{k'+1},
\end{align}
and
\begin{align}
{\tt{T_4}}&=M^2\sigma^2_{e,k}\sigma^2_{e,k+1}\sum_{k'=1}^{K}\sigma^2_{k'}\sigma^2_{k'+1}.\label{eq:t4}
\end{align}

By using \eqref{eq:D_E_kk}, and substituting \eqref{eq:t1}, \eqref{eq:t2}, \eqref{eq:t3} and \eqref{eq:t4} into \eqref{eq:var}, we arrive at the desired result as in \eqref{eq:var2}.

$3)$ Compute ${\tt{IU}}_k$ and ${\tt{AN}}_k$:

Following a similar methodological approach as in 1) and 2), we obtain ${\tt{IU}}_k$ and ${\tt{AN}}_k$  given in \eqref{eq:is} and \eqref{eq:ib1}, respectively.

\section*{Acknowledgment}
This work was supported by project no. 3811/QD-UBND, Binh Duong government, Vietnam. The work of H. Q. Ngo was supported by the Swedish Research Council (VR) and ELLIIT. The work of M. Matthaiou was supported in part by the EPSRC under grant EP/P000673/1.


\begin{thebibliography}{2}
	
	\bibitem{Gunduz:13:IT}
	D. G{\"u}nd{\"u}z, A. Yener, A. Goldsmith, and H. V. Poor, ``The multiway relay channel,'' \emph{IEEE Trans. Inf. Theory}, vol. 59, no. 1, pp. 51-63, Jan. 2013.
	
	\bibitem{AK:09:PIRM}
	A. Amah and A. Klein, ``Non-regenerative multi-way relaying with linear beamforming,'' in \emph{Proc. IEEE PIMRC}, Sept. 2009, pp. 1843-1847.	
	


	\bibitem{ALP:15:TWCOM}
	G. Amarasuriya, E. G. Larsson, and H. V. Poor, ``Wireless information and power transfer in multi-way massive MIMO relay networks,'' \emph{IEEE Trans. Wireless Commun.}, vol. 15, no. 6, pp. 3837-3855, June 2015.
	
	\bibitem{AP:14:PIRM}
	G. Amarasuriya and H. V. Poor, ``Multi-way amplify-and-forward relay networks with massive MIMO,'' in \emph{Proc. IEEE PIMRC}, Sept. 2014, pp. 595-600.
	

	
	\bibitem{ATA:13:TCOM}
	G. Amarasuriya, C. Tellambura, and M. Ardakani, ``Multi-way MIMO amplify-and-forward relay networks with zero-forcing transmission,'' \emph{IEEE Trans. Commun.}, vol. 61, no. 12, pp. 847-4863, Dec. 2013.
	
		\bibitem{NLM:13:TCOM}
		H. Q. Ngo, E. G. Larsson, and T. L. Marzetta, ``Energy and spectral efficiency of very large multiuser MIMO systems,'' \emph{IEEE Trans. Commun.}, vol. 61, no. 4, pp. 1436-1449, Apr. 2013.
		
		\bibitem{Rusek:2013:SPM}
		F. Rusek, D. Persson, B. K. Lau, E. G. Larsson, T. L. Marzetta, O. Edfors, and F. Tufvesson, ``Scaling up MIMO: Opportunities and challenges with very large arrays,'' \emph{IEEE Signal Process. Mag.,} vol. 30, no. 1, pp.~40-60, Jan. 2013.
		
		
		\bibitem{EGL:14:CM}
		E. G. Larsson, O. Edfors, F. Tufvesson, T. L. Marzetta, ``Massive MIMO for next generation wireless systems,'' \emph{IEEE Commun. Mag.}, vol. 52, no. 2, pp. 186-195, Feb. 2014.		
	
	\bibitem{NSML:14:JSAC}
	H. Q. Ngo, H. A Suraweera, M. Matthaiou, and E. G. Larsson, ``Multipair full-duplex relaying with massive arrays and linear processing,'' \emph{IEEE J. Sel. Areas  Commun.}, vol. 32, no. 9, pp. 1721-1737, Sept. 2014.	
	
	
	
		
		
		
		\bibitem{BMM:15:WC}
		E. Bj\"{o}rnson, M. Matthaiou, and M. Debbah, ``Massive MIMO with non-ideal arbitrary arrays: Hardware scaling laws and circuit-aware design,'' \emph{IEEE Trans. Commun.}, vol. 14, no. 8, pp. 4353-4368, Aug. 2015.																													
		
		\bibitem{AMT:04:CIT}
		A. M. Tulino and S. Verd\'{u}, ``Randdom matrix theory and wireless communication,'' \emph{Foundations and Trends in Communications and Information Theory}, vol. 1, no. 1, pp. 1-182, Jun. 2004.
		
		\bibitem{Kay:93:ECNJ}
		S. M. Kay, \emph{Fundamentals of Statistical Signal Processing: Estimation Theory}. Englewood Cliffs, NJ: Prentice Hall, 1993.			
		
		\bibitem{ngo2014massive}
		H. Q. Ngo, M. Matthaiou, and E. G. Larsson, ``Massive MIMO with optimal power and training duration allocation,'' \emph{IEEE Wireless Commun. Lett.}, vol. 3, no. 6, pp. 605-608, Sept. 2014.
		
			
		\bibitem{KZMEZ:16:CIT}
		C. Kong, C. Zhong, M. Matthaiou, E. Bj\"{o}rnson, and Z. Zhang, ``Multipair two-way half-duplex relaying with massive arrays and imperfect CSI,'' \emph{IEEE Trans. Inf. Theory}, Jul. 2016, submitted. [Online]. Available: https://arxiv.org/abs/1607.01598.
		
%
\end{thebibliography}

\end{document}